\documentclass[12pt]{article}
\usepackage{amsmath}

\usepackage{amssymb}

\usepackage{euscript}

\usepackage{epsfig}  
  
\usepackage{cite}  
  
\usepackage{alltt}  
\def \bea{\begin{eqnarray}}
\def \beq{\begin{equation}}

\def \to {\rightarrow}
\def \eea{\end{eqnarray}}
\def \eeq{\end{equation}}

\begin{document}
\begin{titlepage}
%Finding the CP Nature of the Higgs Boson at the NLC.
\begin{flushright}  
{\bf   
IFT/02-xxx \\
May 2002}  
\end{flushright}  

\hspace{1 mm}  
\vspace{1 cm}  
\begin{center}{\bf\Large {\boldmath  Measuring Higgs Coupling to 
{\boldmath$\tau$}-Leptons at }}
\end{center}  
\begin{center}{\bf\Large  Future {\boldmath$e^+ e^-$} Colliders} \end{center}   
\vspace{0.3 cm}  
\begin{center}  
{\large\bf A. Chalov$^{a}$ ,~ A. Likhoded$^{a,b,}
$\footnote{andre@ift.unesp.br}  ~ and ~  R. Rosenfeld$^{b,}$
\footnote{rosenfel@ift.unesp.br}}  
\vspace{0.3 cm}  
\\  
{\em $^a$ Institute of High Energy Physics, \\
Protvino, Moscow Region, Russia}\\  
{\em $^b$ Instituto de F\'{\i}sica Te\'orica - UNESP \\
 Rua Pamplona, 145, 01405-900 S\~ao Paulo, SP, Brasil}\\    
\vspace{0.5 cm}  
  
\end{center}

\centerline {\bf Abstract}
\vskip 1.0cm

We perform a complete simulation of the process $ e^+ e^- \to \tau^+ \tau^- \nu
\bar{\nu}$, where $\nu$ can be an electron, muon or tau neutrino, 
in the context of a general Higgs coupling to $\tau$ leptons. 
We analyse various kinematical distributions and
obtain the sensitivity regions in the parameter space that can be explored at 
a future  $e^+ e^-$ collider. In particular, inclusion of $W$ boson fusion 
enhances the sensitivity significantly.

\bigskip

\vfill
\end{titlepage}

\newpage

\section{Introduction}

\indent
The origin of fermion masses and mixings is one of most important issues in
particle physics. Unfortunately, these parameters are inputs in the well-tested 
Standard Model (SM). All one can do is to measure them as accurately as 
possible and hope that in the future a more fundamental theory will 
actually be able to predict their values. 

Fermion masses are possibly related to the electroweak symmetry 
breaking mechanism, which is not known at the
moment. In the simplest model, a scalar electroweak doublet in an {\it ad hoc}
quartic
potential is responsible for the symmetry breaking, leaving a scalar physical
boson, the Higgs boson (J$^{\mbox{{\tiny PC}}} = 0^{++}$), as a remnant. 
Favorite extensions of the SM, like  the minimal supersymmetric standard model
 (MSSM),
also predict the existence of a heavy pseudoscalar boson 
(J$^{\mbox{{\tiny PC}}} = 0^{-+}$), in addition to a 
light scalar boson. Another possibility is that the electroweak
symmetry breaking is triggered by some new strong interactions, and in this case
the lightest boson could be a pseudoscalar, 
like a pion \cite{techni}. Therefore, it is very
important to distinguish between these two cases. 
This can be achieved by studying the parity properties of the scalar particle, 
once it is found. The general framework for performing this study in the decay
of the Higgs into fermion or gauge boson pairs was developed in \cite{early} and
applied to Higgs production via Higgsstrahlung $e^+e^- \to Z H$ in
\cite{higgstrahlung}.

Possibly the most direct way to measure the CP properties of the Higgs boson is
in the environment of a photon-photon collider, by setting the polarization of
the photon in order to select different CP states \cite{photons}.

Details of the $ZHH$ couplings can also be measured in a model independent
way by studying the threshold behaviour and the angular distribution of the 
Higgsstrahlung process. In particular, the spin and parity assignements of the
Higgs boson can be determined \cite{zhh}.

At future hadron machines, like the CERN Large Hadron Collider (LHC), weak gauge 
boson fusion can provide an efficient way to measure the CP properties of the
$HWW$ coupling by analysing the azimuthal angle distribution of the two 
outgoing forward tagging jets \cite{LHC}. 

The coupling of the Higgs boson to top quarks can also be used as a measure of
the CP properties of the higgs boson in the processes $pp \rightarrow t \bar{t}
h$ at the LHC \cite{topLHC} and $ e^+ e^- \rightarrow t \bar{t} H$ at a future
Linear Collider \cite{topLC}.
Prospects of measuring the parity of the Higgs boson have also been analysed in 
$\mu^+ \mu^-$  machines \cite{muon}.

In this paper we concentrate on the determination of the (pseudo)scalar-tau-tau 
coupling at a Linear Collider with a center-of-mass energy of $\sqrt{s} = 500$
GeV and accumulated luminosity of 1 ab$^{-1}$, based on the Tesla proposal
 \cite{Tesla}. We wil assume
that this particle has already been discovered at the Large Hadron Collider but
a detailed study of its couplings is missing. 

Contrary to previous works \cite{tau,bower}, where only Higgsstrahlung and Higgs
radiation off a heavy fermion leg were 
considered, we take into account all relevant contributions to the process 
$ e^+ e^- \to \tau^+ \tau^- \nu
\bar{\nu}$, where $\nu$ can be an electron, muon or tau neutrino. In
particular, weak gauge boson fusion is the dominant contribution to Higgs
production for $M_H < 180$ GeV at $\sqrt{s} \ge 500$ GeV.

In extensions of the SM with extra scalars and pseudoscalars, the lightest
spin-0 particle can be an admixture of states without a definite parity
\cite{tau}. Hence, 
we parametrize the general $H \tau \tau$ coupling as:
\begin{equation}
\frac{m_\tau}{v} (a + i \gamma_5 b) 
\end{equation}
where $v = 246$ GeV and $a = 1, b = 0$  in the Standard Model. We will present 
results 
considering $a$ and $b$ as independent parameters and also for the cases of 
fixed $a=1$, free $b$ and fixed $b=0$, free $a$. We'll see that there is a
region of insensivity around circles of $\sqrt{a^2+b^2}=1$ in the $a-b$ plane.

Without considering  $\tau$ decays, we are only sensitive to terms 
proportional to
$a$ , which comes from the interference with non-Higgs contributions, and
$a^2 $ and $b^2$ from pure Higgs contributions. Therefore, even
at the $\tau$ level we can search for deviations from the Standard
Model prediction which could arise, for instance, in supersymmetric models.

The analysis of $\tau$ decays permits to study some P-odd correlations which
isolates the effects of the $b$ term and hence can distinguish between a
scalar and pseudoscalar (or a CP-violating mixture) nature of the Higgs-tau-tau
coupling. This will be the subject of a forthcoming publication.

\section{Monte-Carlo simulation}

\begin{figure}
\centerline{\epsfxsize=1\hsize \epsffile{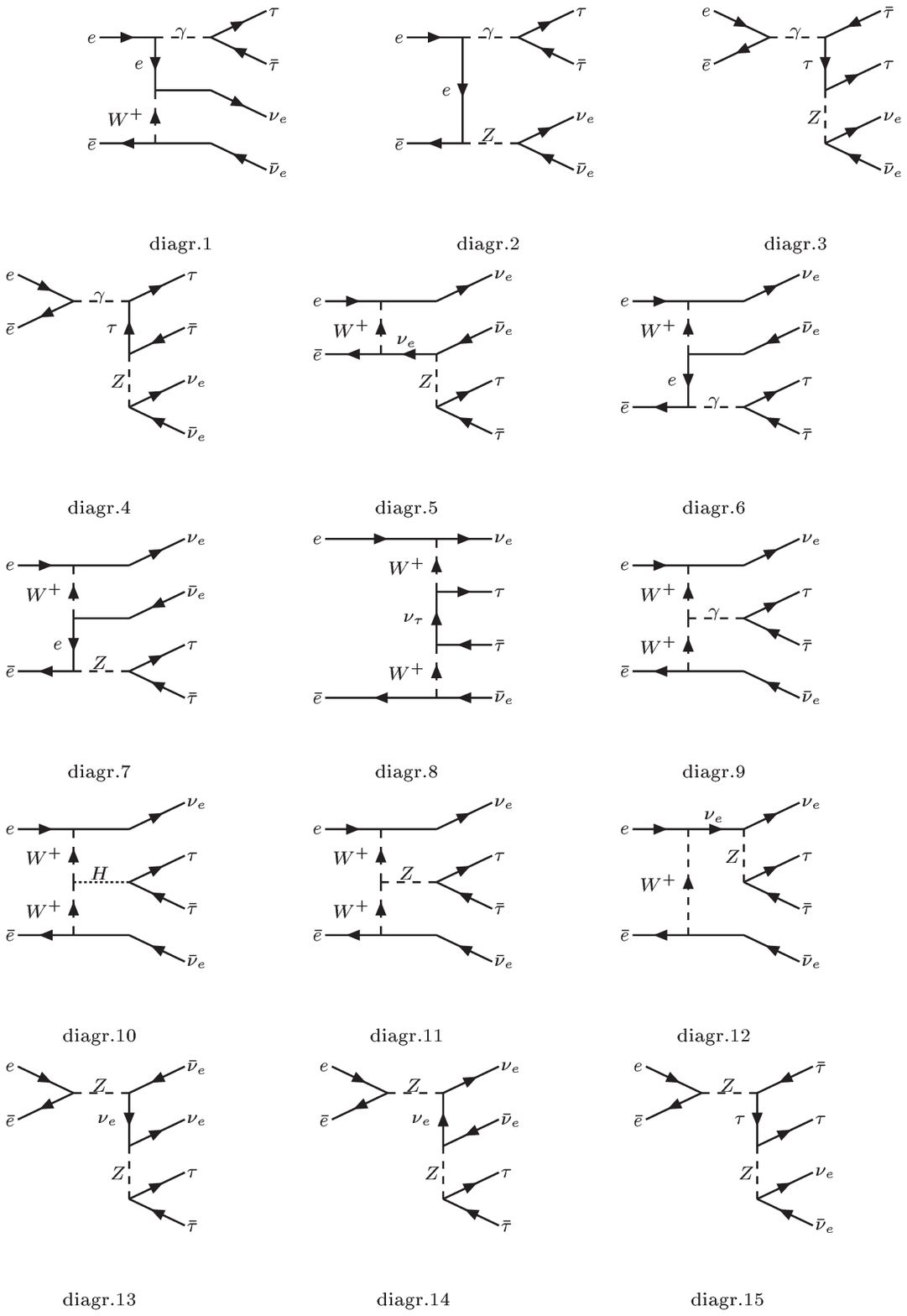}
\hskip-7cm \epsfxsize=1\hsize \epsffile{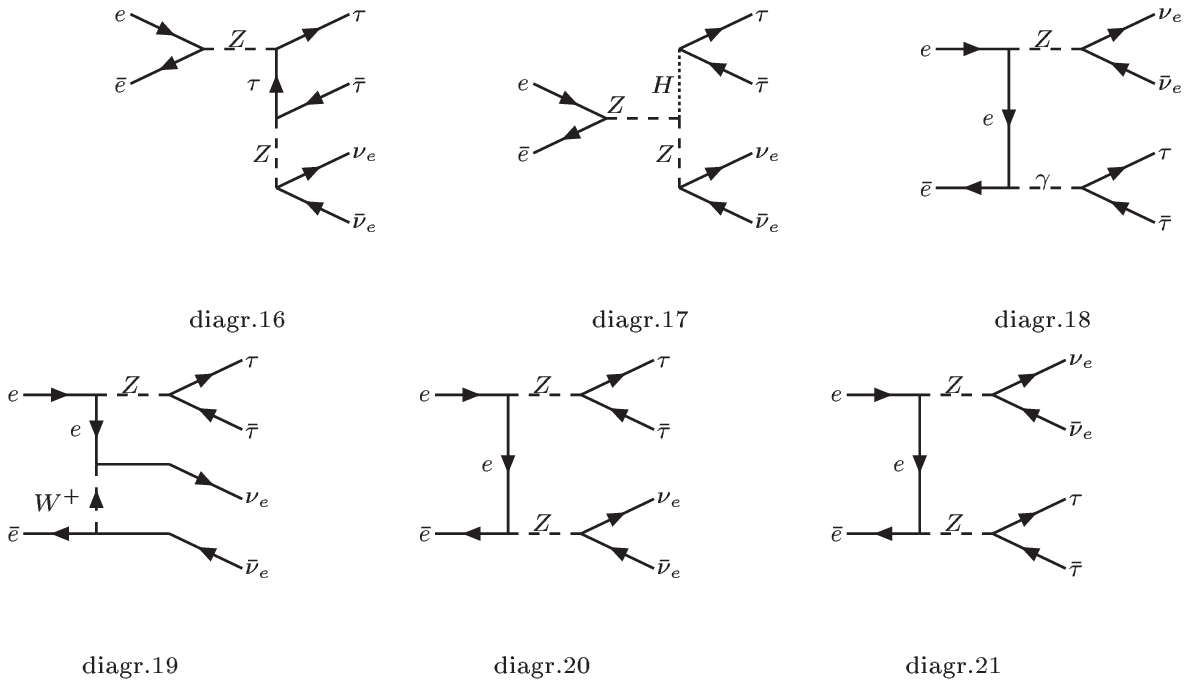}}
\vskip-2cm
\caption{}
\label{fig1}
\end{figure}

We generate all possible contributions to the processe $ e^+ e^- \to 
\tau^+ \tau^- \nu \bar{\nu}$, where $\nu$ can be either an electron, muon or tau
neutrino.
The following contributions are computed:

$\bullet$ $e^+e^-\to \tau^+\tau^- \nu_e\bar\nu_e$: there are 21 Feynman diagrams
that contribute to this process (see Fig. \ref{fig1}), of which only 
Diag. 10 (W-fusion) and Diag. 17 (Higgssthralung) are signal contributions;

\begin{figure}
\centerline{\epsfxsize=1\hsize \epsffile{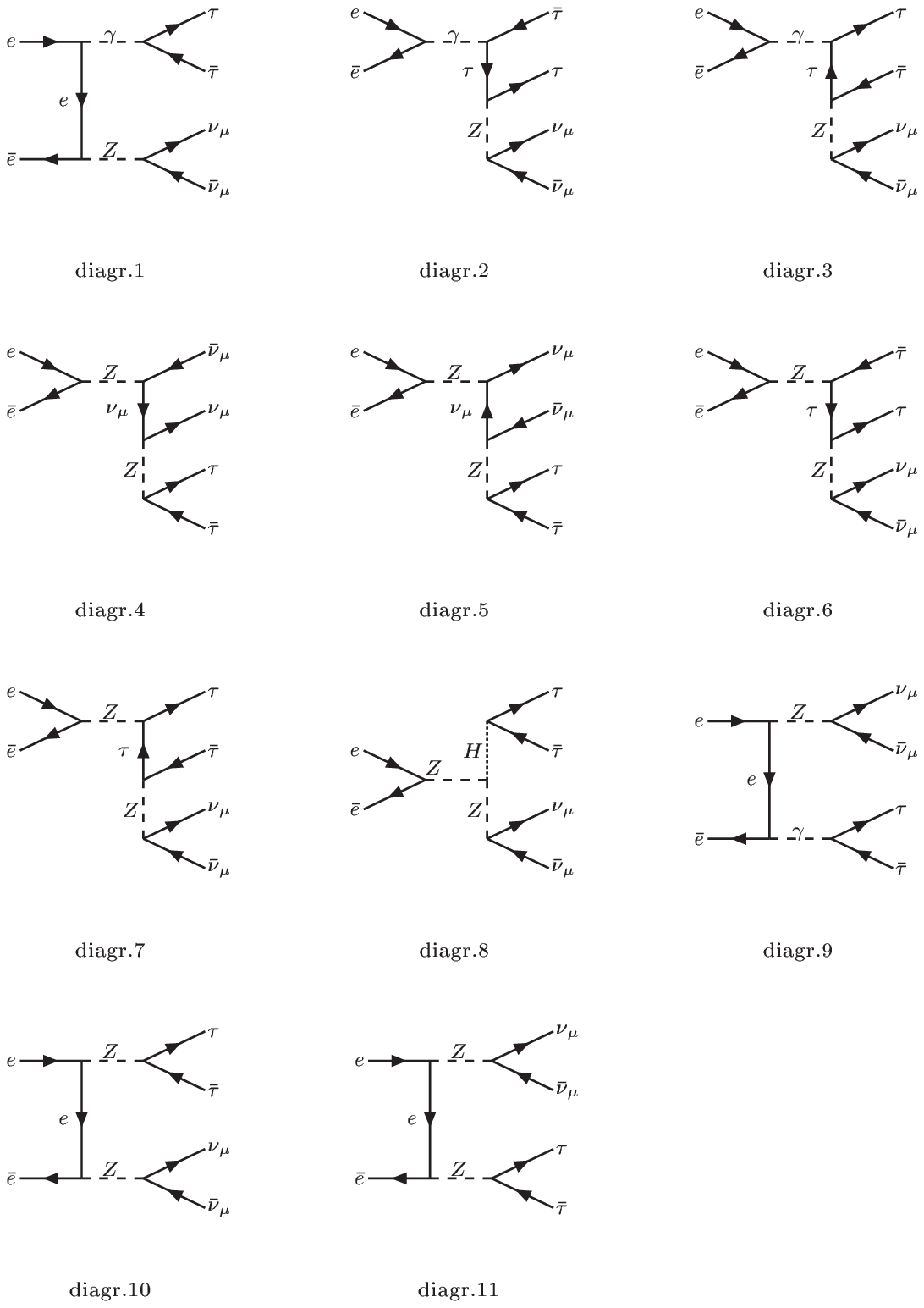}}
\vskip-2cm
\caption{}
\label{fig2}
\end{figure}

$\bullet$ $e^+e^-\to \tau^+\tau^- \nu_\mu\bar\nu_\mu$: there are 11 Feynman
diagrams in this case (see Fig. \ref{fig2}), of which only Diag. 8 
(Higgssthralung) contributes to the signal;

\begin{figure}
\centerline{\epsfxsize=1\hsize \epsffile{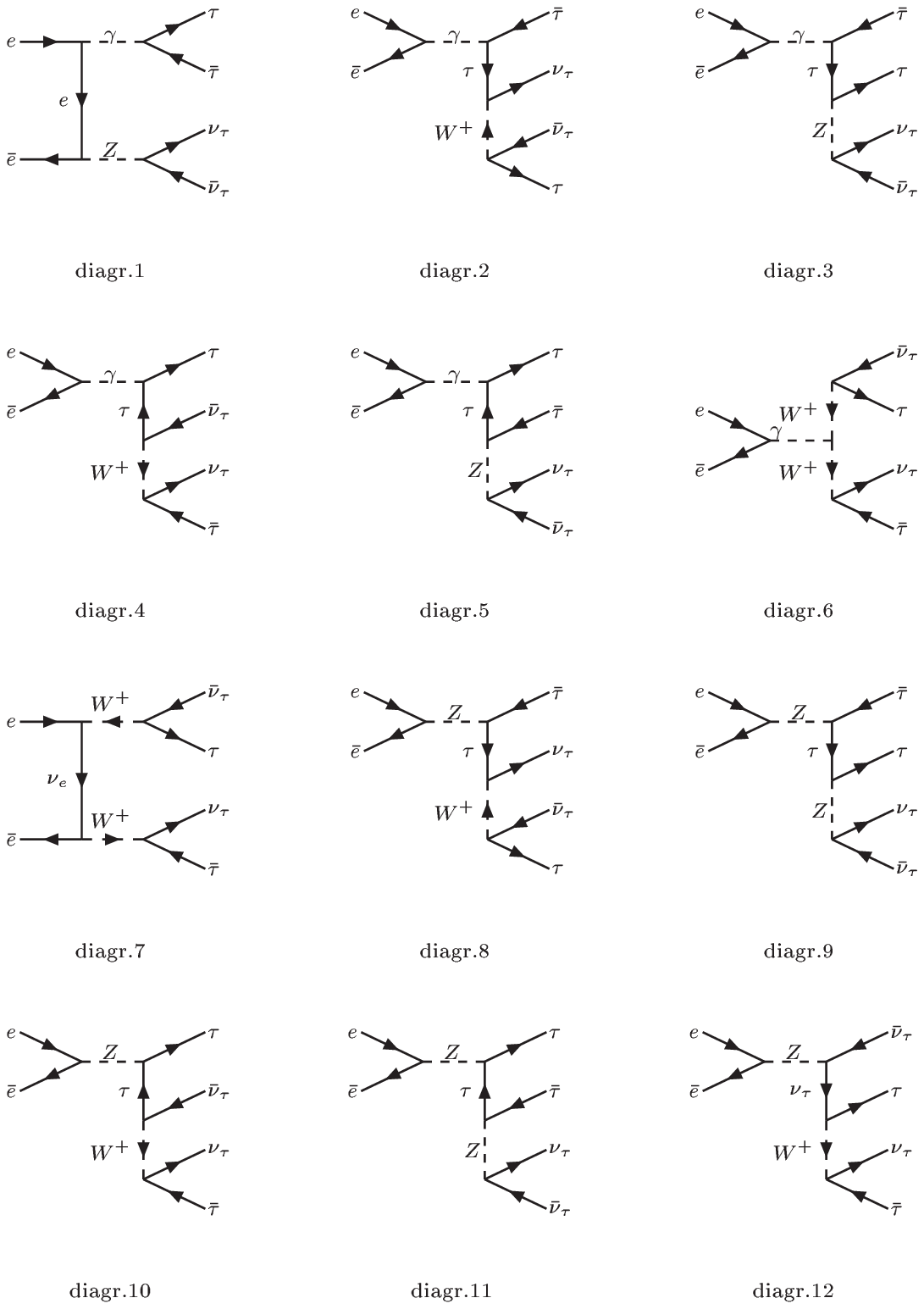}
\hskip-7cm \epsfxsize=1\hsize \epsffile{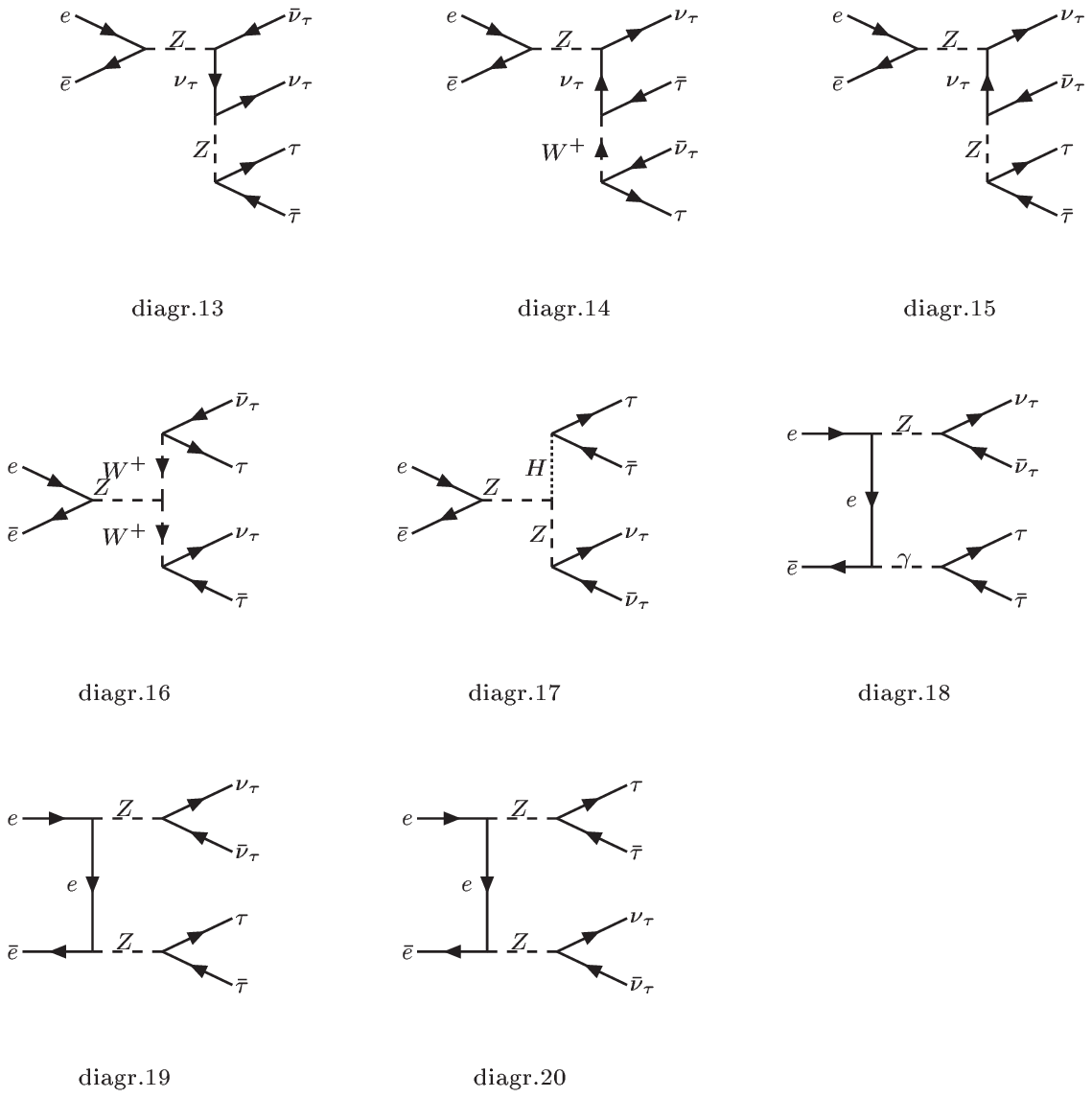}}
\vskip-2cm
\caption{
}
\label{fig3}
\end{figure}

$\bullet$ $e^+e^-\to \tau^+\tau^-\nu_\tau\bar\nu_\tau $: there are 20 Feynman
diagrams in this case (see Fig. \ref{fig3}), of which only Diag. 17 
(Higgssthralung) contributes to the signal.

The cross sections for these three contributions, with $m_H = 100$ GeV for
illustration, are respectively 
$\sigma_{\nu_\tau} = 2.81 \times 10^{-2}$ pb, 
$\sigma_{\nu_\mu} = 1.25 \times 10^{-2}$ pb, and
$\sigma_{\nu_e} = 0.109$ pb.
As expected,  $\sigma_{\nu_e}$ is dominant due to the fusion diagram.

\begin{figure}
\centerline{\epsfxsize=1\hsize \epsffile{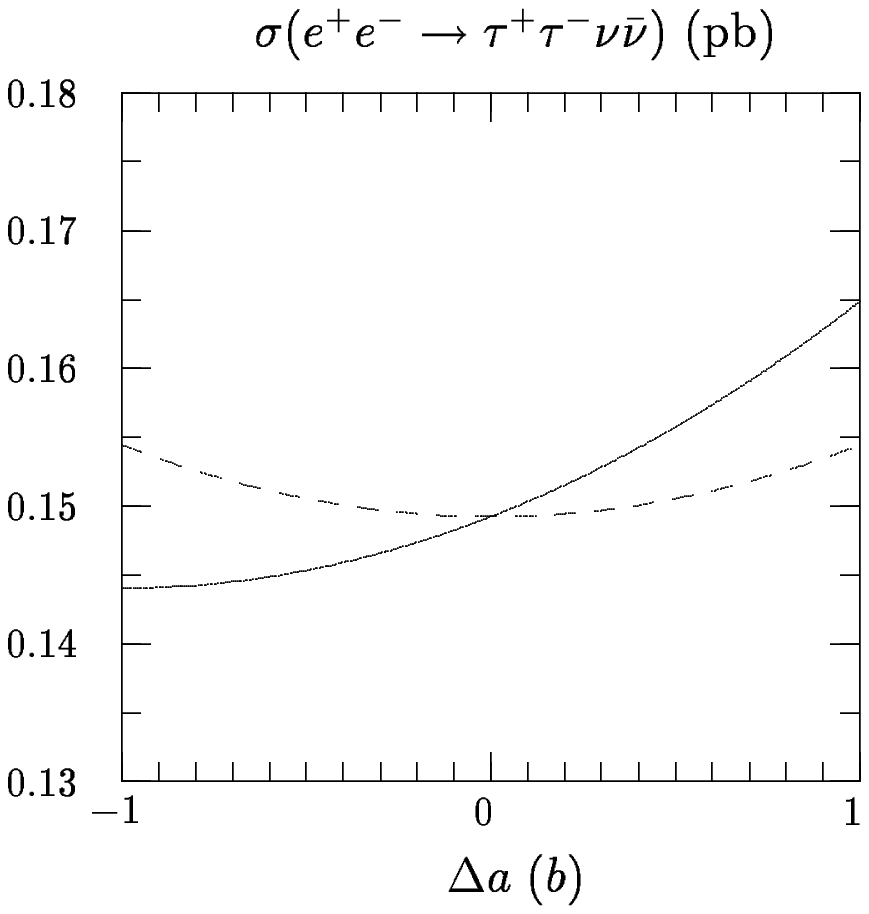}}
\vskip-10cm
\caption{}
\label{cross}
\end{figure}

In Fig. \ref{cross} we show the dependence of the total cross section 
$e^+e^-\to \tau^+\tau^-\nu \bar\nu $, summed over the three neutrino species, on
the parameters $\Delta a = a - 1$ and $b$. The dependence in $b$ is symmetric
since only terms proportional to $b^2$ contribute to the cross section. On the
other hand, the dependence on $\Delta a$ is asymmetric due to interference with
standard model processes, which leads to the presence of linear terms in 
$\Delta a$. Our goal is to see how sensitive the experiments
performed at the next generation of $e^+e^-$ colliders will be in the 
determination of these parameters. 

\section{Analysis}

%%%%%%%%%%%%%%%%%%%%%%%%%%%%%%%%%%%%%%%%%%%%%%%%%%%
\subsection{Experimental Aspects }

\noindent
To proceed with the analysis of the $e^+e^-\to \tau^+\tau^-\nu\bar\nu$ 
process one needs to have the  final state topology as well as the $\tau$'s 
momenta reconstructed.

For the generated data sample of the $e^+e^-\to \tau^+\tau^-\nu\bar\nu$
process we assume that tau-lepton pairs in the final state
can be reconstructed in the experiment by using $\tau$ decays into $\pi$ and
$\rho$. The $\tau\to\pi\nu$ and $\tau\to \rho \nu$ decay channels are the two most 
important and together provide about 13\% of the total branching ratio of $\tau$
pairs. As it was shown in the LEP/SLC studies, other channels, such as
$\tau^\pm\to \pi^-\pi^+\pi^\pm\nu$, $\tau^\pm\to \pi^0\pi^0\pi^\pm\nu$, as well
as leptonic tau decay modes, can be included increasing the experimental
data sample to 82\% .
 
 There are several methods  which can be used to reconstruct the final 
$\tau$ pair topology \cite{reco1}. 
Recently, a new method \cite{reco2}
was proposed, which allows one to have a high reconstruction efficiency in the 
$\tau^\pm\to \rho^\pm \bar\nu_\tau (\nu_\tau )$, $\rho^\pm\to \pi^\pm\pi^0$
decay chain. This method was used to study the Higgs boson's parity using 
$\tau\to\rho\nu$ \cite{parity}.
  
The results of the LEP experiments \cite{LEP} show that the $\tau$ 
reconstruction 
efficiency varies from 51\%  to 80\% .
Based on the above results we adopt the moderate $\tau$-pair
reconstruction efficiency  $\varepsilon_{\tau \tau} = 50$\% .

Another important aspect is the assumption about the detector performance 
and possible sources of the systematic uncertaintes. We 
include the anticipated  systematic errors of 0.5\% in the luminosity 
measurement,
1\%  in the acceptance determination, 1\% in the branching ratios, and
1\% in the background substraction, and assume the Gaussian nature of the 
sytematics. To place bounds on the $H\tau^+\tau^-$ couplings, we  use a standard 
$\chi^2$-criterion to analyse the events. We found that the 
most strict bounds are achieved by dividing the distribution event samples 
into 10 bins.

As for the possible contribution from background processes, like
$e^+e^-\to e^+e^- ZZ\to e^+e^- \tau^+\tau^-\nu\bar\nu$ (with $e$-pair lost),
$e^+e^-\to \nu\bar\nu W^+W^- \to \nu\bar\nu \tau^+\tau^-\nu\bar\nu$, 
$e^+e^-\to ZZZ\to \tau^+\tau^-\nu\bar\nu\nu\bar\nu$ etc., the cross-sections
of these processes are either small, or they can be significantly
supressed without much difficulties, as it was shown in the analogous study
for the process $e^+e^-\to\nu\bar\nu H\to \nu\bar\nu b\bar b$ \cite{LC}.

%%%%%%%%%%%%%%%%%%%%%%%%%%%%%%%%%%%%%%%%
\subsection{Kinematical Distributions}

\begin{figure}
\centerline{\epsfxsize=0.8\hsize \epsffile{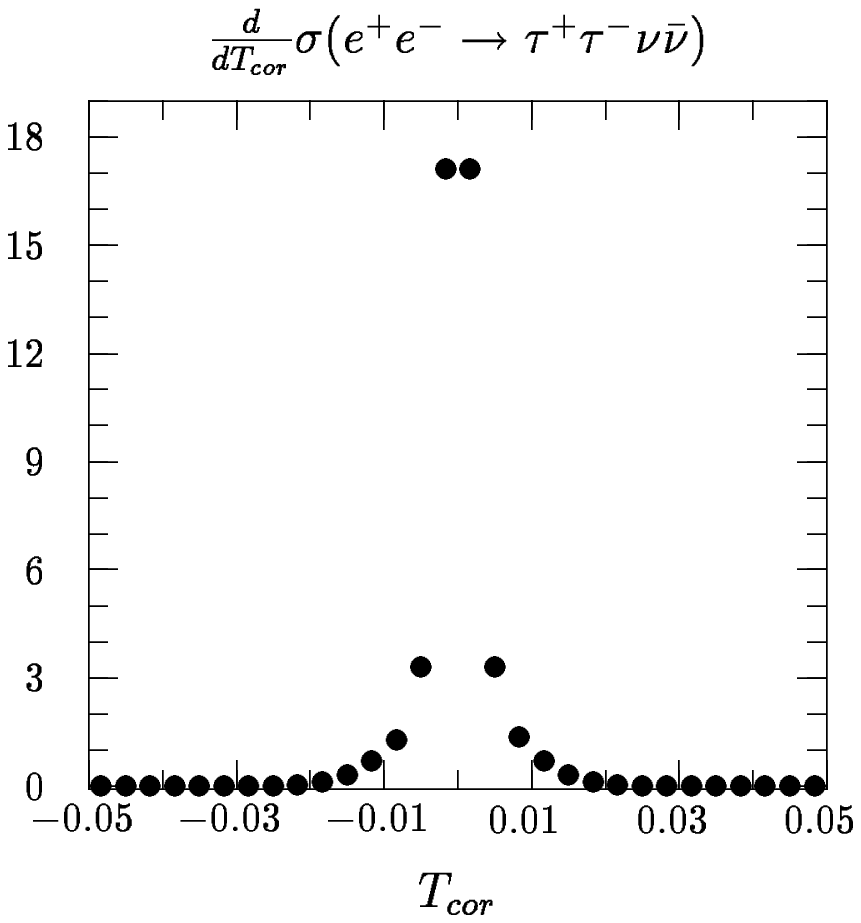}
\hskip-5cm \epsfxsize=0.8\hsize \epsffile{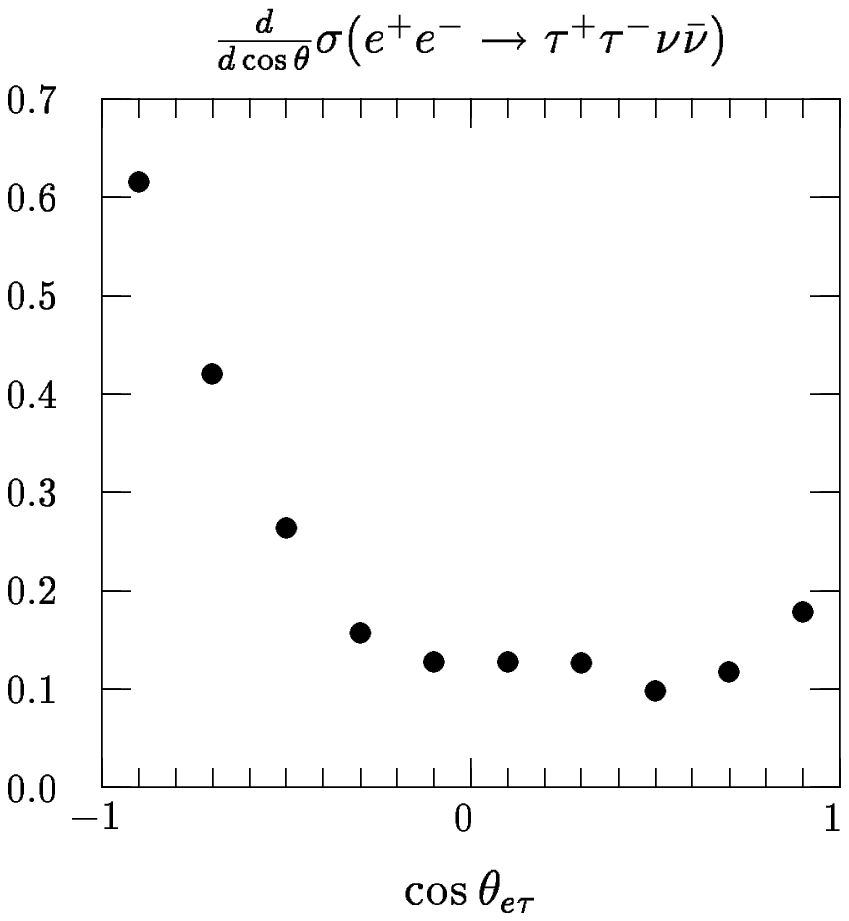}}
\vskip-10cm
\centerline{\epsfxsize=0.8\hsize \epsffile{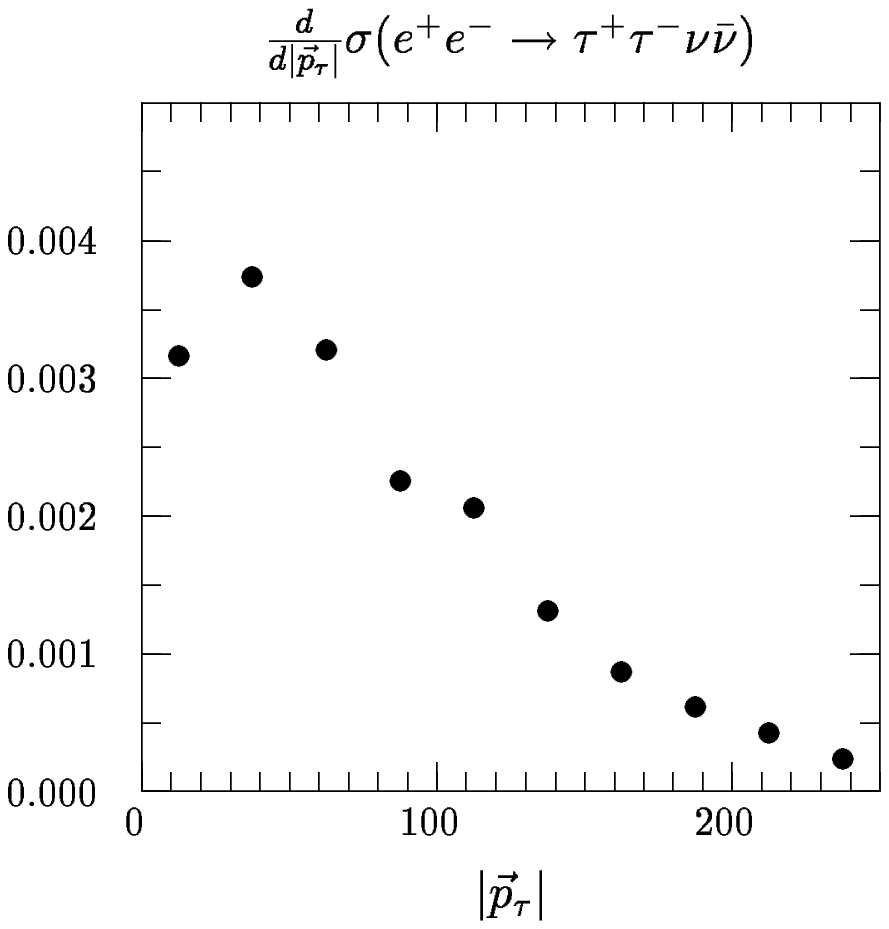}
\hskip-5cm \epsfxsize=0.8\hsize \epsffile{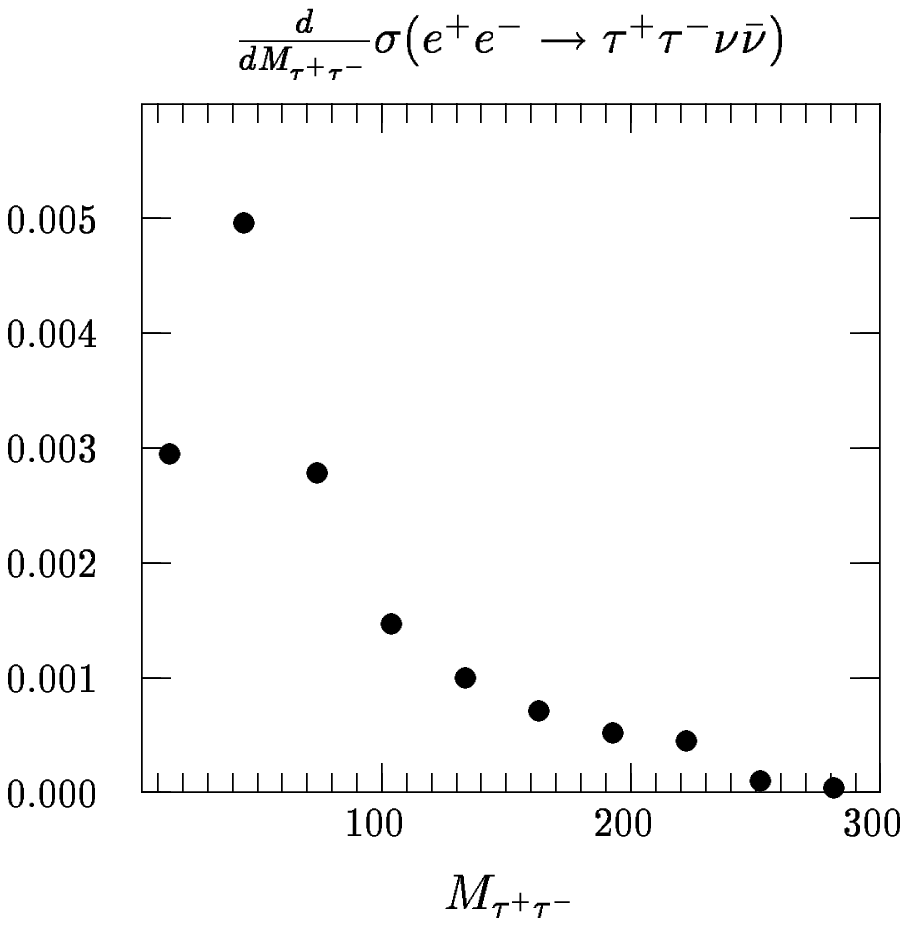}}
\vskip-9cm
\caption{}
\label{distributions}
\end{figure}

We analysed various distributions in order to determine which one is most
sensitive to new physics: in the $T$ correlation, 
defined by $T = \frac{1}{(\sqrt{s})^3}
\vec{p}_{el} \cdot    (\vec{p}_{\tau_1} \times \vec{p}_{\tau_2})$, the $\tau$ 
momentum $|\vec{p}_{\tau}|$, the angle between $\tau$ and the beam  
$\cos \theta_{e \tau}$ and the $\tau$ pair invariant mass $M_{\tau^+ \tau^-}$.
They are shown in Fig. \ref{distributions} for the Standard Model, 
all plotted with 10 bins for illustration purposes.

\begin{figure}
\centerline{\epsfxsize=0.8\hsize \epsffile{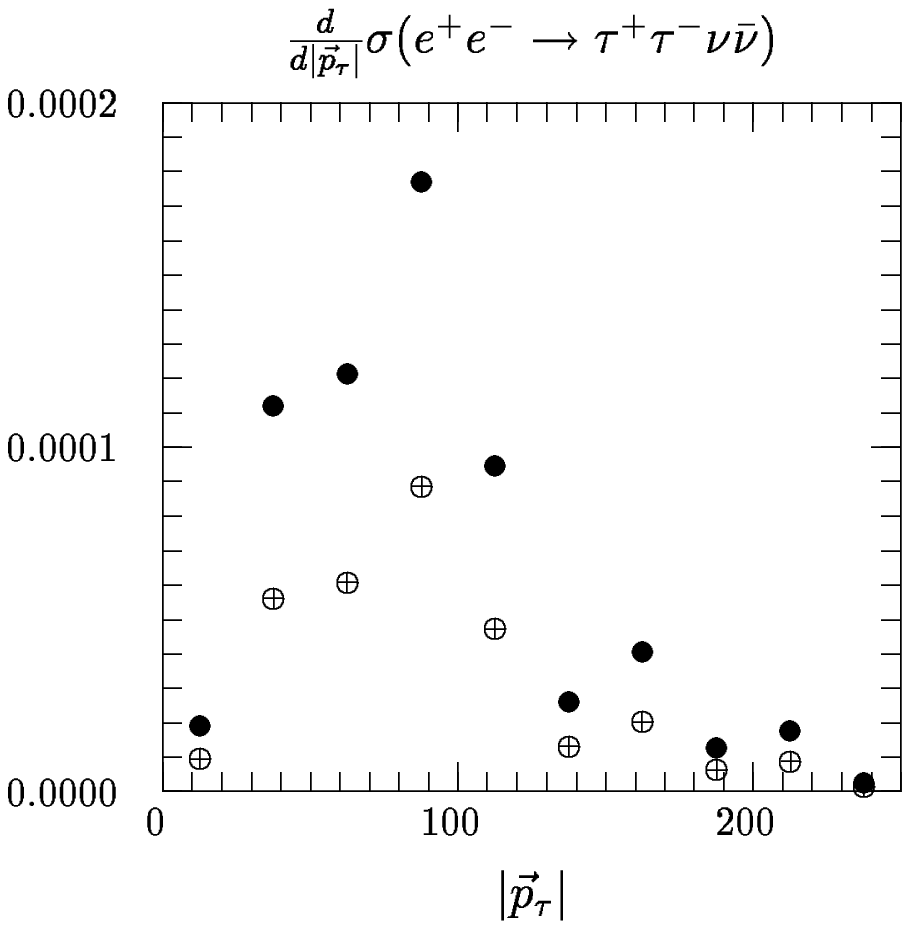}
\hskip-5cm \epsfxsize=0.8\hsize \epsffile{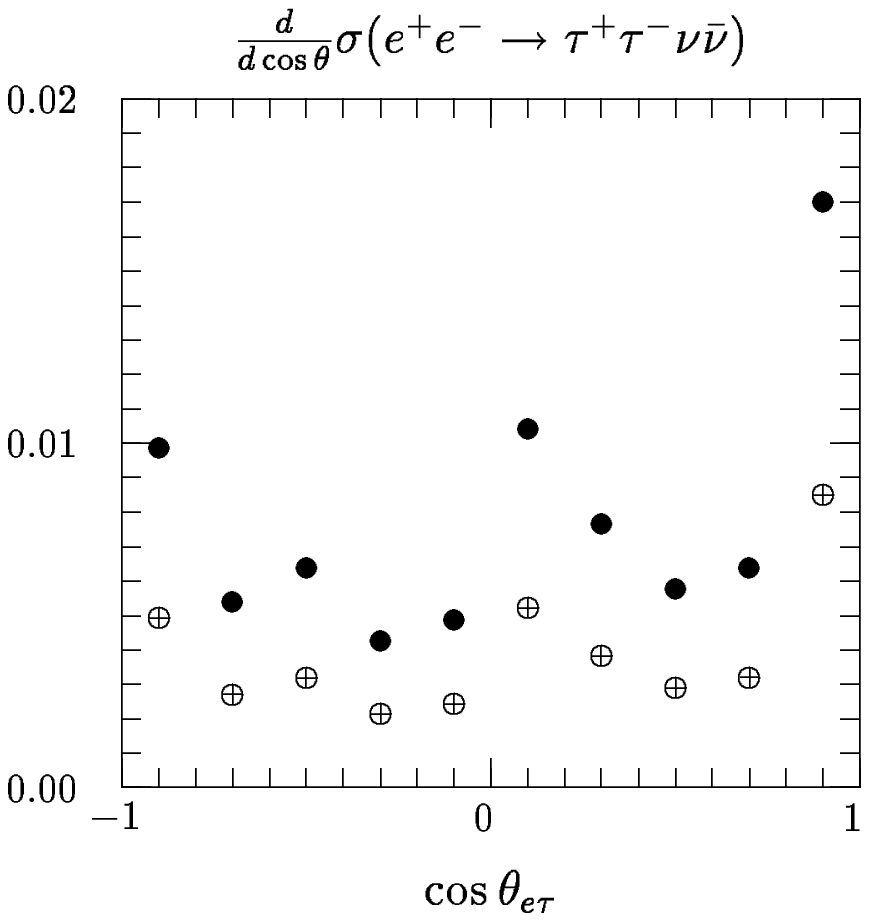}}
\vskip-10cm
\centerline{\epsfxsize=0.8\hsize \epsffile{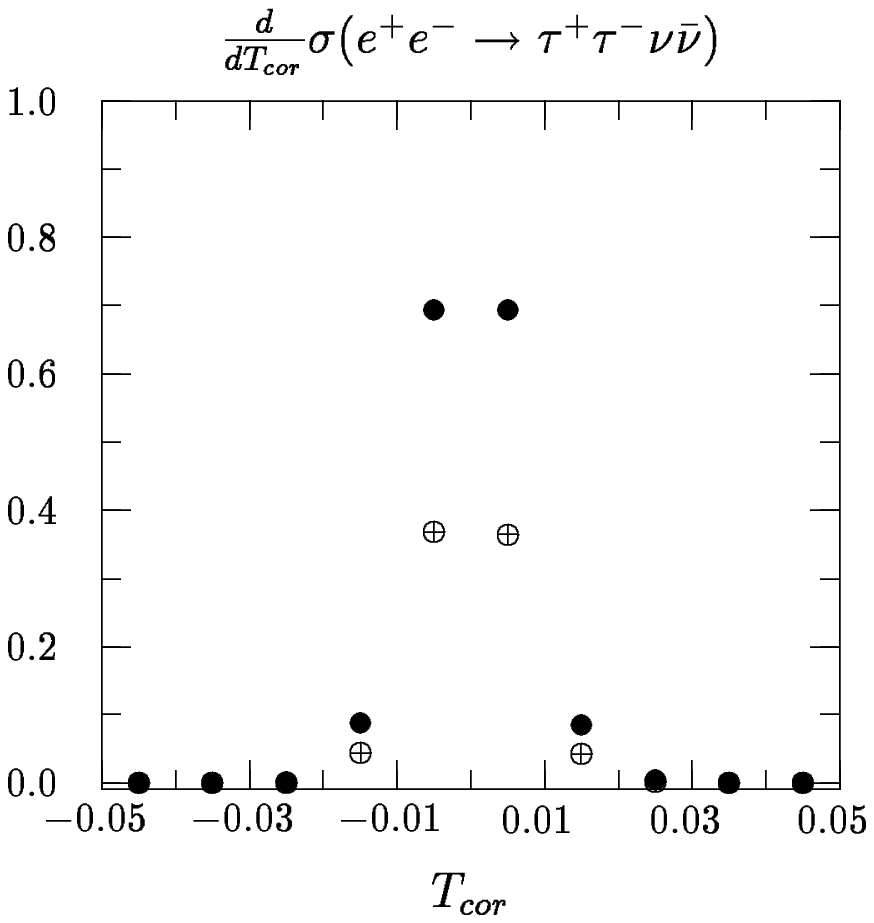}
\hskip-5cm \epsfxsize=0.8\hsize \epsffile{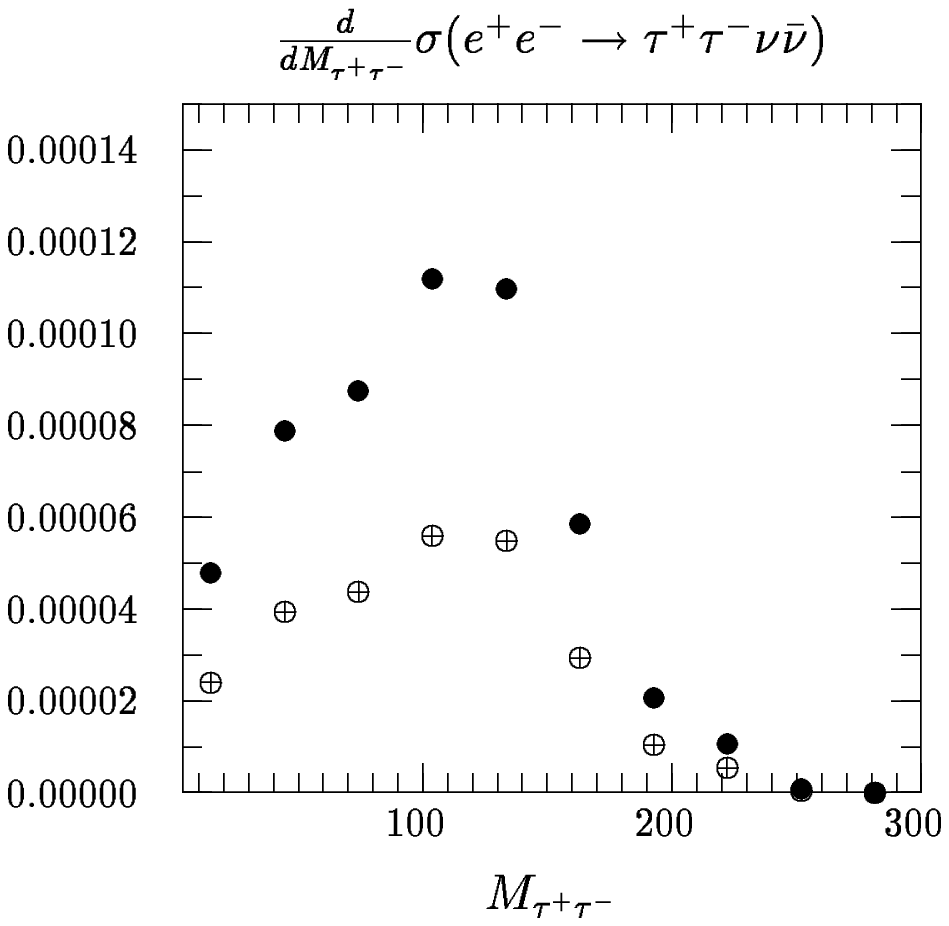}}
\vskip-9cm
\caption{}
\label{higgsonly}
\end{figure}

In Fig.\ref{higgsonly}, we illustrate the difference in the Higgs 
contributions in these distributions for the standard model ($a=1, b=0$) and
($a=0.5, b=0.5$). We see that the shapes are very similar, as expected, but the
levels can be noticeable different.

\begin{figure}
\centerline{\epsfxsize=0.8\hsize \epsffile{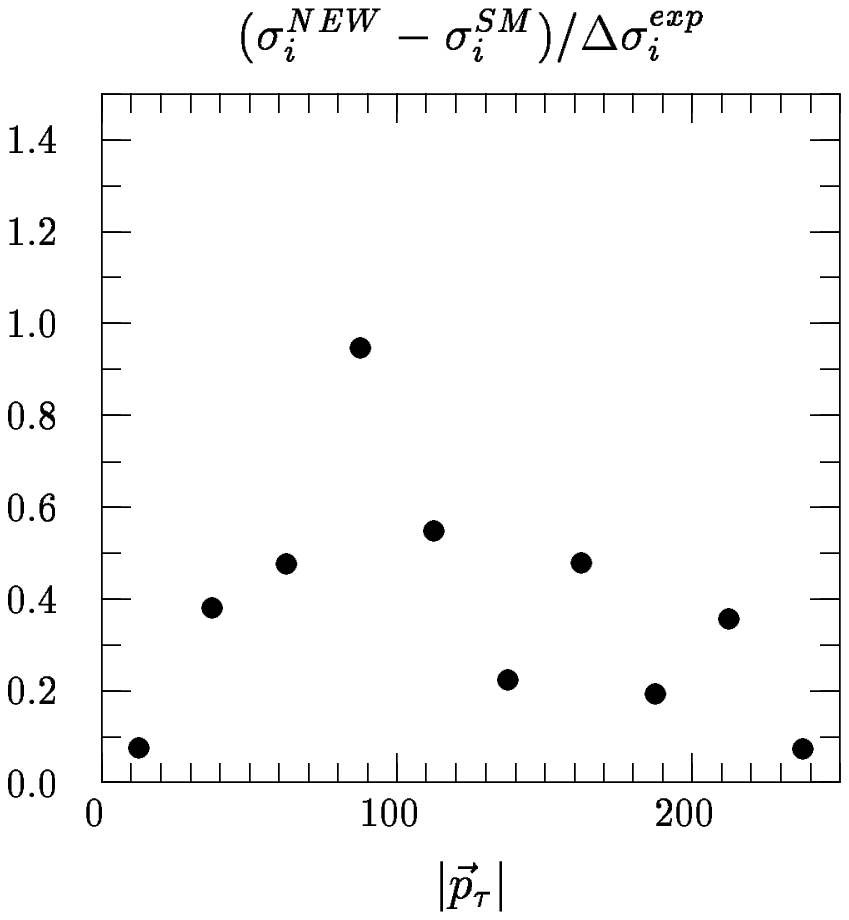}
\hskip-5cm \epsfxsize=0.8\hsize \epsffile{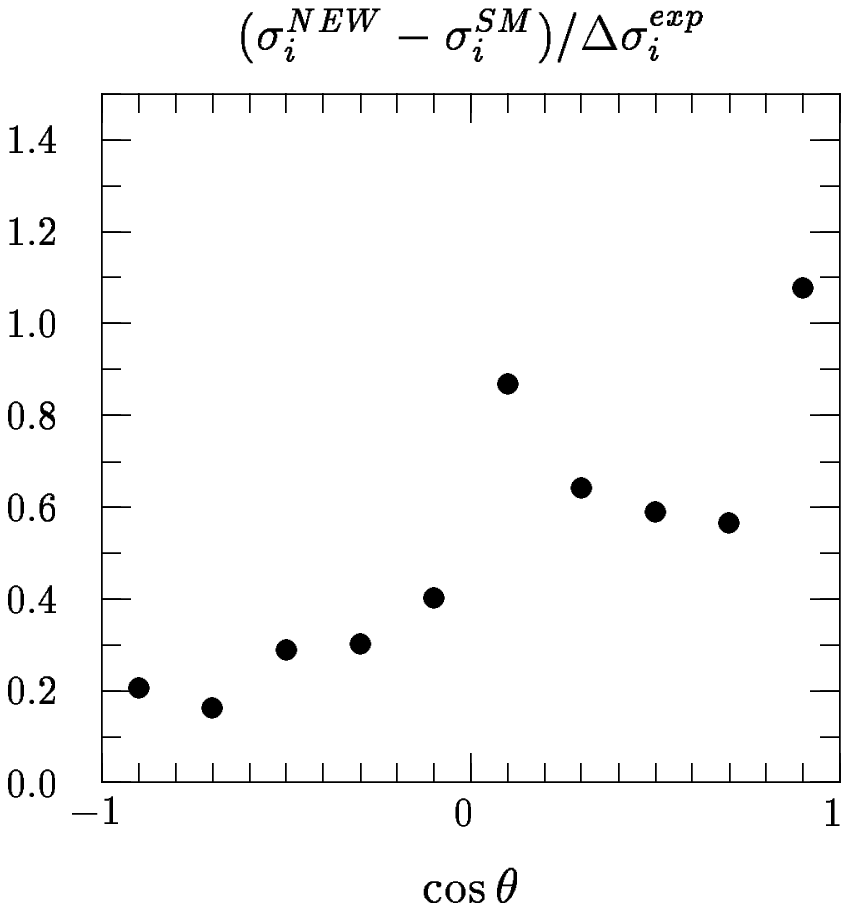}}
\vskip-10cm
\centerline{\epsfxsize=0.8\hsize \epsffile{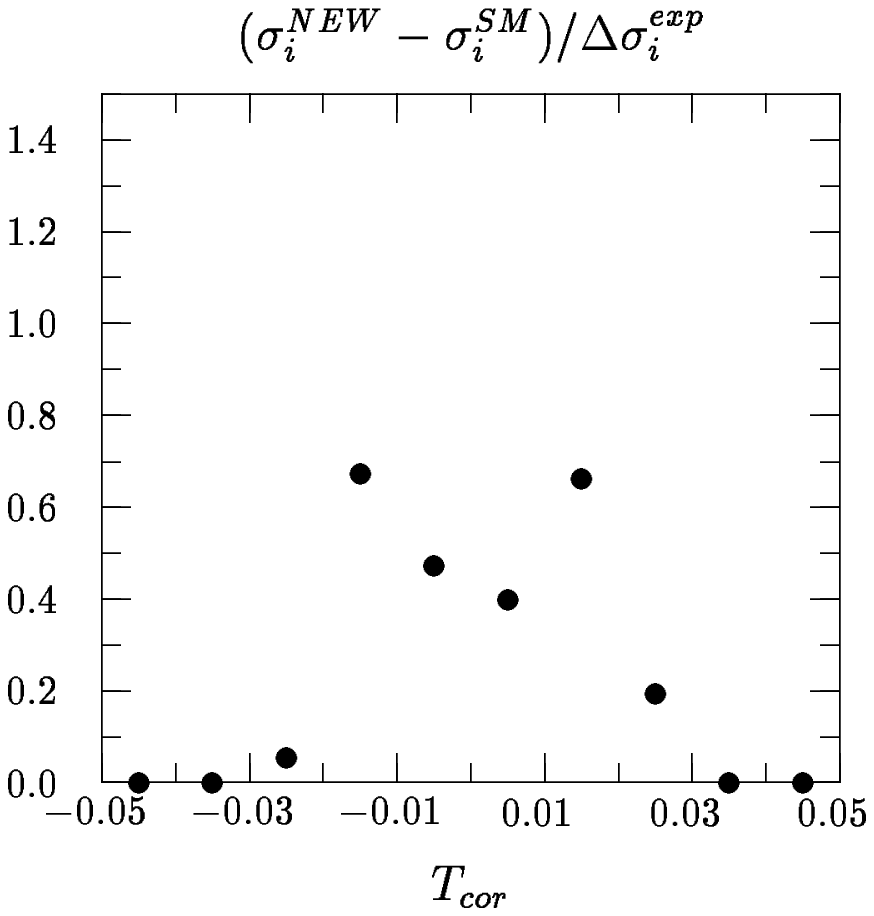}
\hskip-5cm \epsfxsize=0.8\hsize \epsffile{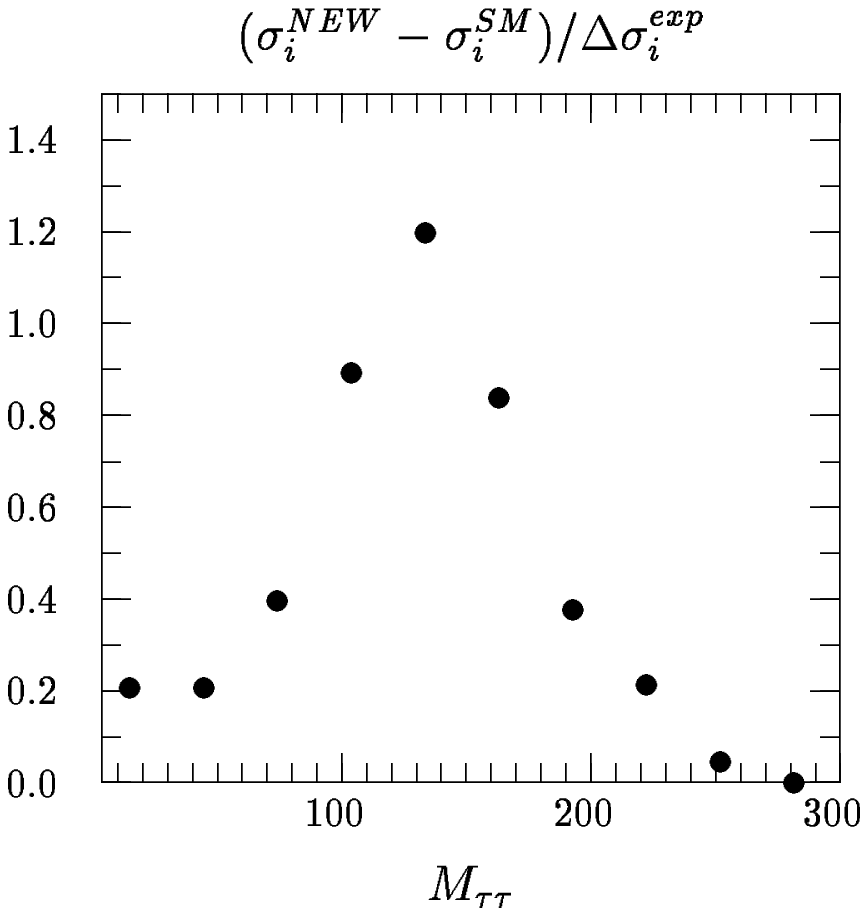}}
\vskip-9cm
\caption{}
\label{TotalSensitivity}
\end{figure}

In order to find the level of sensitivity, we show in 
Fig.\ref{TotalSensitivity} the quantity 
$\frac{(\sigma^{New}_i - \sigma^{SM}_i)}{\Delta \sigma^{exp}_i}$ for each bin of the
distributions, where $\sigma^{New}$ is the cross-section with
$a$ and/or $b$ different from their SM values. In this figure, we used 
$a=1, b=0.5$ as an example of new  physics. A luminosity  
$\int {\cal L} dt = 1$ ab$^{-1}$ was used. We would like to stress again that
including the contribution from $W$ boson fusion enhances significantly the 
sensitivity and the result seems to be not so pessimistic as pointed out 
in \cite{bower}.

In order to determine the sensitivity regions in the $a-b$ plane 
that can be probed in the 
next generation of linear colliders, we used the standard $\chi^2$ method where
the experimental error $\Delta
\sigma^{exp}_i$ is given by:
\begin{equation}
\Delta\sigma^{exp}_i = \sigma^{SM}_i \sqrt{\delta^2_{syst} + \delta^2_{stat} }
\end{equation}
where 
\begin{equation}
\delta_{stat}  = \frac{1}{\sqrt{\sigma^{SM}_i \varepsilon_{\tau \tau} \int {\cal L} dt 
}}
\end{equation}
and $\delta_{syst}^2$ is the sum in quadrature of the systematic uncertainties
mentioned above.

\begin{figure}
\centerline{\epsfxsize=0.8\hsize \epsffile{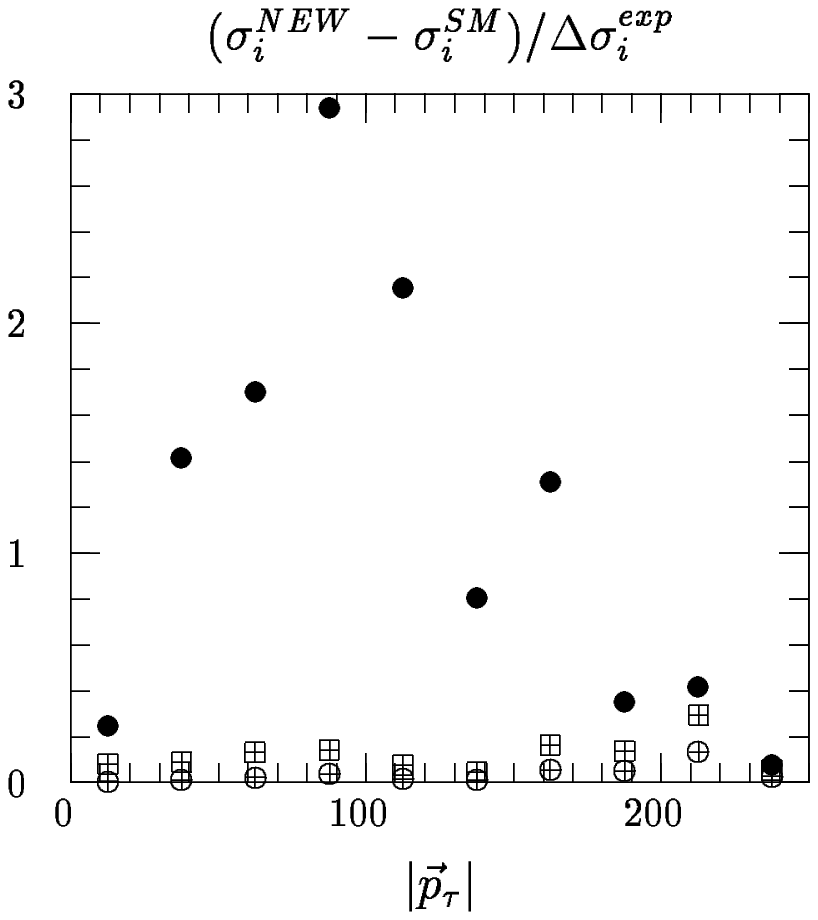}
\hskip-5cm \epsfxsize=0.8\hsize \epsffile{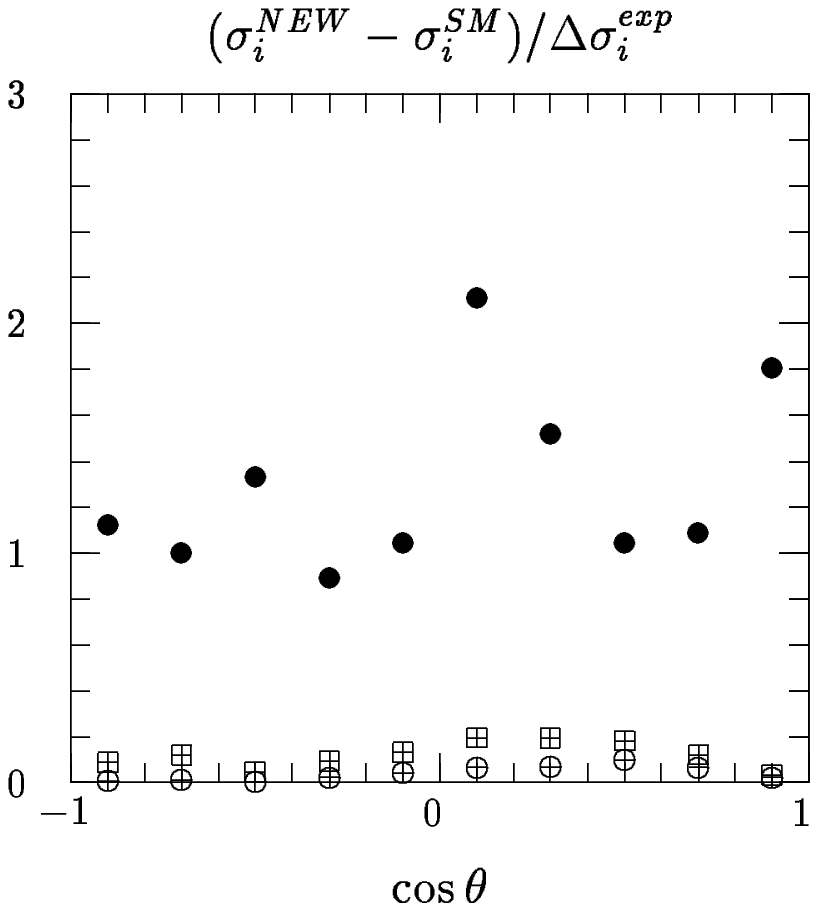}}
\vskip-10cm
\centerline{\epsfxsize=0.8\hsize \epsffile{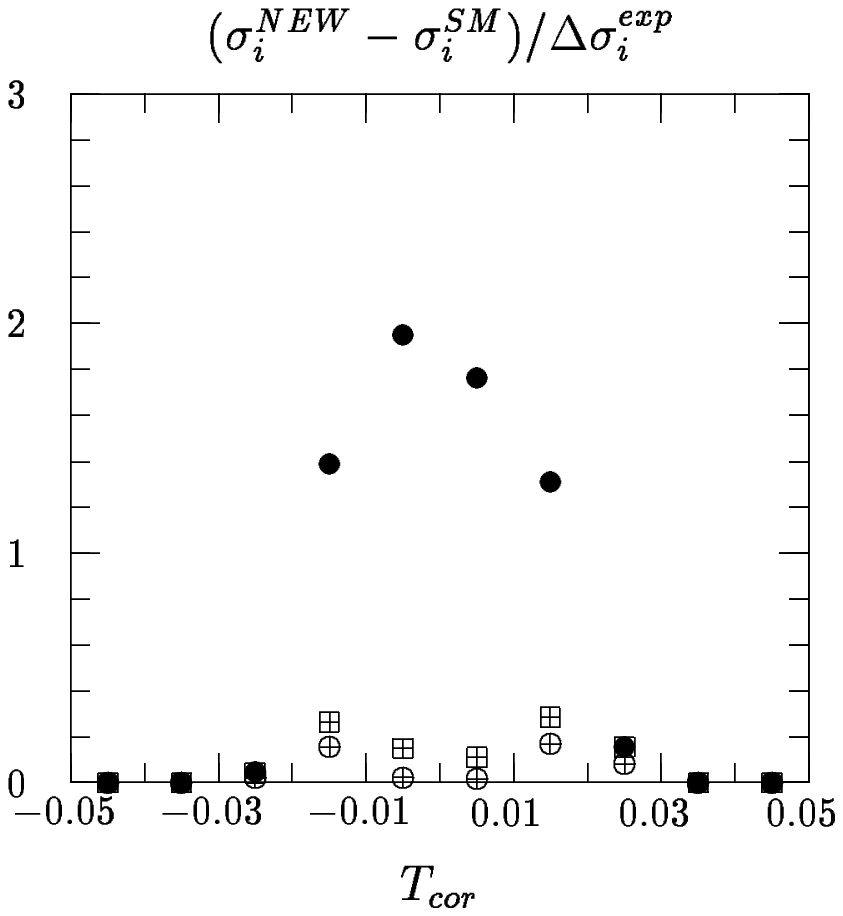}
\hskip-5cm \epsfxsize=0.8\hsize \epsffile{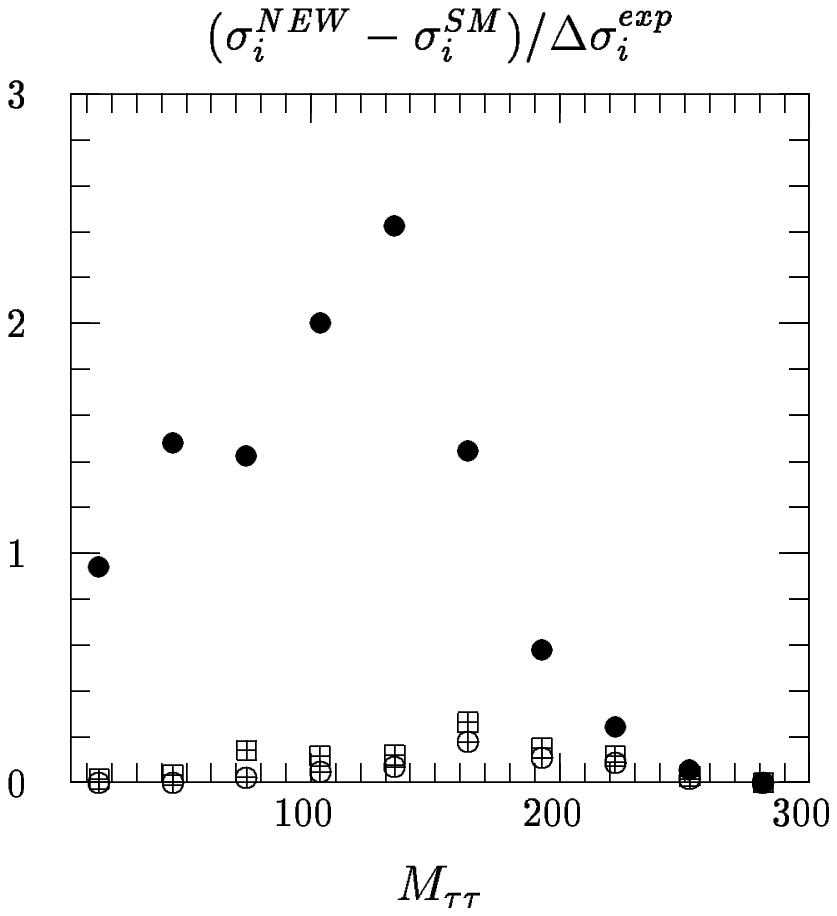}}
\vskip-9cm
\caption{} 
\label{relative}
\end{figure}

In order to understand the relative contributions for each class of diagramas in
Figs. 1--3, we show in Fig.\ref{relative} the individual contributions from
electron neutrinos, muon neutrinos and tau neutrinos channels. As expected, 
electron neutrinos dominate due to the weak gauge boson fusion contribution.

\section{Results}

In order to determine the sensitive region in the parameter space, we decided to
use the $\cos \theta_{e \tau}$ distribution, since the $\tau$ direction is
possibly the easiest observable to be reconstructed from data. We also found 
that using 10 bins in the analysis optimizes the results.

\begin{figure}
\centerline{\epsfxsize=1\hsize \epsffile{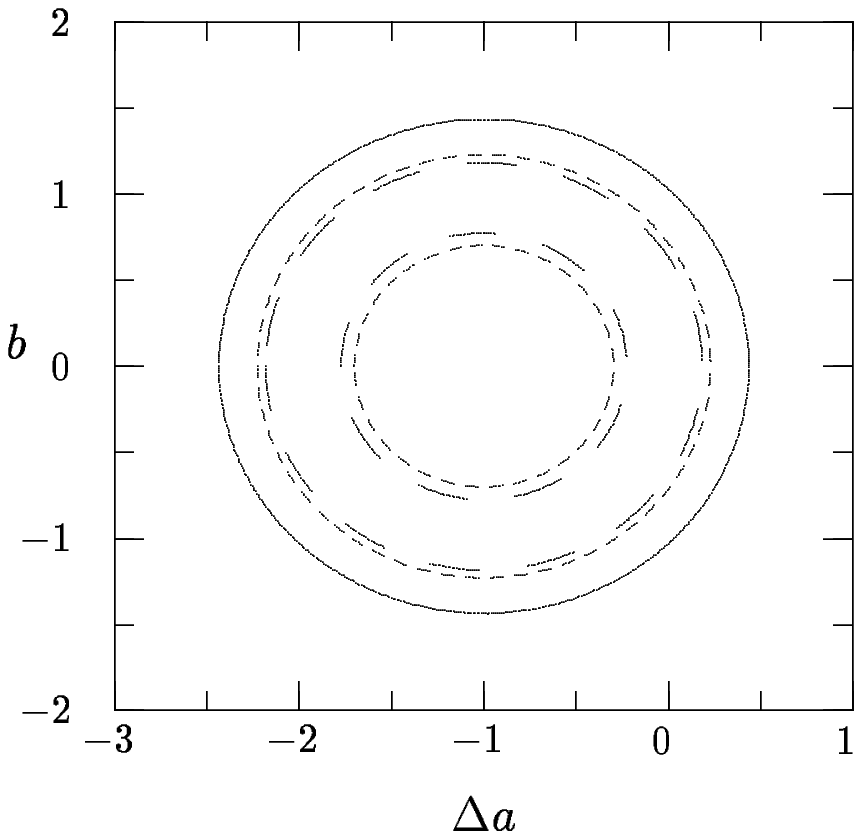}}
\vskip-10cm
\caption{ 
}
\label{luminosity}
\end{figure}

We investigated the effects of the luminosity and the results are presented
in Fig. \ref{luminosity}, where the sensitivity regions 
for $\int {\cal L} dt =$ 100 fb$^{-1}$, 1 ab$^{-1}$ and 10 ab$^{-1}$ are shown. 
We can easily see what can be gained from an increase of
luminosity at TESLA.

\begin{figure}
\centerline{\epsfxsize=1\hsize \epsffile{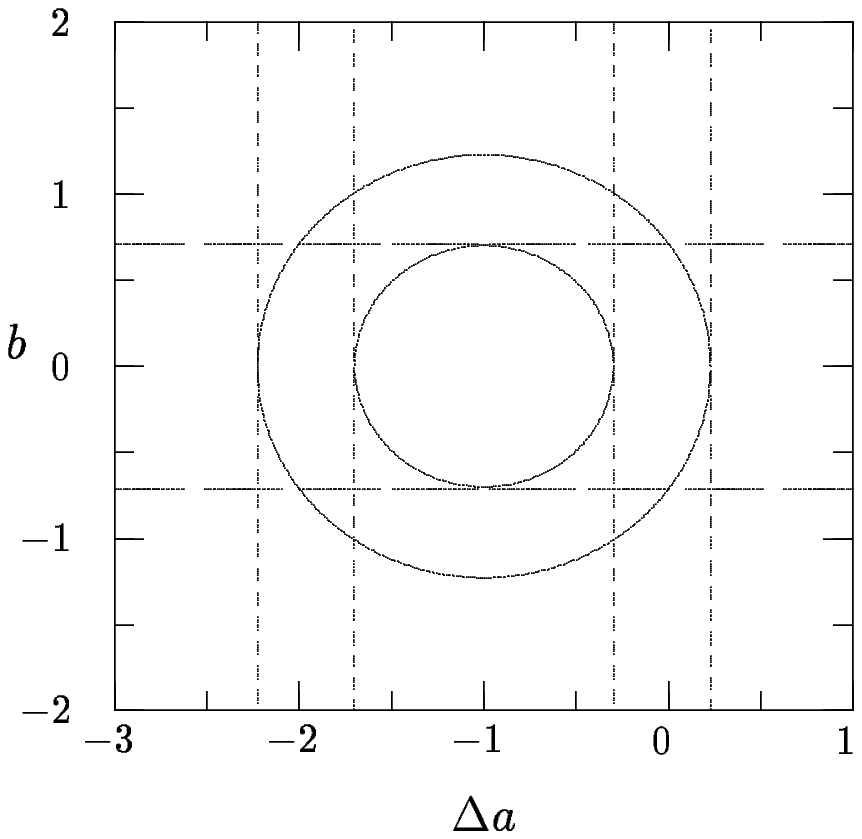}}
\vskip-10cm
\caption{   
}
\label{contour_final}
\end{figure}

In Fig. \ref{contour_final}, we show our final results for a Tesla-like
environment, with luminosity of 1 ab$^{-1}$ and center-of-mass energy of 500 GeV
and for $M_H= 100$ GeV.
We adopted a 50 \% $\tau$-pair reconstruction efficiency and a systematic
error of $\simeq 2$\% , in accordance with Tesla Design Report \cite{Tesla}. 
The allowed region for independent $\Delta a$ and $b$ parameters 
at 95\% confidence level is the area between the solid circles.
The horizontal band with long-dashed lines is the allowed region for the $b$
parameter keeping $a=1$. The two vertical bands with short-dashed lines are the 
the allowed region for the $\Delta a$
parameter keeping $b=0$ (of course only the right band containing the SM point
is physical).

The bounds that can be obtained at 95\% confidence level  are:
\begin{description}
\item[1.] The case of two independent parameters: 
\begin{displaymath}
(0.70)^2 \leq (\Delta a+1)^2+b^2 \leq (1.23)^2\;.
\end{displaymath}
\item[2.] The case of $b=0$ and free $\Delta a$: 
\begin{displaymath}
        -0.30 \leq \Delta a\leq 0.23\;.
\end{displaymath}
\item[3.] The case of $\Delta a=1$ and free $b$:
\begin{displaymath}
       -0.7 \leq b\leq 0.7\;.
\end{displaymath}
\end{description}

These results can be easily scaled for moderate variations in the Higgs boson 
mass around $100$ GeV by multiplying the bounds by a factor $(M_H/100 \mbox{
GeV})^2$.

\section{Conclusions}

We have performed a  complete analysis of the sensitivity to new $h \tau \tau$ 
couplings from the process $ e^+ e^- \to \tau^+ \tau^- \nu
\bar{\nu}$ at the next generation of linear colliders. These new couplings are
predicted by many extensions of the Standard Model. We showed that forthcoming
experiments will be able to probe deviations of $H \tau \tau$ coupling.
For a Tesla-like environment, we are able to constrain the couplings in the
region
$(0.70)^2 \leq (\Delta a+1)^2+b^2 \leq (1.23)^2$ for independent $a$ and $b$
parameters. In order to disentangle the individual contributions of the 
parameters $a$ and $b$, and to determine the $CP$ nature of the Higgs boson, 
further analysis of $\tau$ decay products is needed. 
This will be the subject of a 
separate publication.

\section*{Acknowledgments}

The work of A. Likhoded was partially funded by a Fapesp grant 
2001/06391-4. The work of A. Chalov is partially supported
by the Russian Foundation for Basic
Research, grants 99-02-16558 and 00-15-96645, Russian Education Ministry,
grant RF~E00-33-062, and CRDF grant MO-011-0.
R. Rosenfeld would like to thank Fapesp and CNPq for partial
financial support. We would like to thank Claudio Dib for useful comments.

% Journal and other miscellaneous abbreviations for references
\def \arnps#1#2#3{Ann.\ Rev.\ Nucl.\ Part.\ Sci.\ {\bf#1} (#3) #2}
\def \art{and references therein}
\def \cmts#1#2#3{Comments on Nucl.\ Part.\ Phys.\ {\bf#1} (#3) #2}
\def \cn{Collaboration}
\def \cp89{{\it CP Violation,} edited by C. Jarlskog (World Scientific,
Singapore, 1989)}
\def \econf#1#2#3{Electronic Conference Proceedings {\bf#1}, #2 (#3)}
\def \efi{Enrico Fermi Institute Report No.\ }
\def \epjc#1#2#3{Eur.\ Phys.\ J. C {\bf#1} (#3) #2}
\def \f79{{\it Proceedings of the 1979 International Symposium on Lepton and
Photon Interactions at High Energies,} Fermilab, August 23-29, 1979, ed. by
T. B. W. Kirk and H. D. I. Abarbanel (Fermi National Accelerator Laboratory,
Batavia, IL, 1979}
\def \hb87{{\it Proceeding of the 1987 International Symposium on Lepton and
Photon Interactions at High Energies,} Hamburg, 1987, ed. by W. Bartel
and R. R\"uckl (Nucl.\ Phys.\ B, Proc.\ Suppl., vol.\ 3) (North-Holland,
Amsterdam, 1988)}
\def \ib{{\it ibid.}~}
\def \ibj#1#2#3{~{\bf#1} (#3) #2}
\def \ichep72{{\it Proceedings of the XVI International Conference on High
Energy Physics}, Chicago and Batavia, Illinois, Sept. 6 -- 13, 1972,
edited by J. D. Jackson, A. Roberts, and R. Donaldson (Fermilab, Batavia,
IL, 1972)}
\def \ijmpa#1#2#3{Int.\ J.\ Mod.\ Phys.\ A {\bf#1} (#3) #2}
\def \ite{{\it et al.}}
\def \jhep#1#2#3{JHEP {\bf#1} (#3) #2}
\def \jpb#1#2#3{J.\ Phys.\ B {\bf#1} (#3) #2}
\def \jpg#1#2#3{J.\ Phys.\ G {\bf#1} (#3) #2}
\def \mpla#1#2#3{Mod.\ Phys.\ Lett.\ A {\bf#1} (#3) #2}
\def \nat#1#2#3{Nature {\bf#1} (#3) #2}
\def \nc#1#2#3{Nuovo Cim.\ {\bf#1} (#3) #2}
\def \nima#1#2#3{Nucl.\ Instr.\ Meth. A {\bf#1} (#3) #2}
\def \npb#1#2#3{Nucl.\ Phys.\ B {\bf#1} (#3) #2}
\def \npps#1#2#3{Nucl.\ Phys.\ Proc.\ Suppl.\ {\bf#1} (#3) #2}
\def \npbps#1#2#3{Nucl.\ Phys.\ B Proc.\ Suppl.\ {\bf#1} (#3) #2}
\def \PDG{Particle Data Group, D. E. Groom \ite, \epjc{15}{1}{2000}}
\def \pisma#1#2#3#4{Pis'ma Zh.\ Eksp.\ Teor.\ Fiz.\ {\bf#1} (#3) #2 [JETP
Lett.\ {\bf#1} (#3) #4]}
\def \pl#1#2#3{Phys.\ Lett.\ {\bf#1} (#3) #2}
\def \pla#1#2#3{Phys.\ Lett.\ A {\bf#1} (#3) #2}
\def \plb#1#2#3{Phys.\ Lett.\ B {\bf#1} (#3) #2}
\def \pr#1#2#3{Phys.\ Rev.\ {\bf#1} (#3) #2}
\def \prc#1#2#3{Phys.\ Rev.\ C {\bf#1} (#3) #2}
\def \prd#1#2#3{Phys.\ Rev.\ D {\bf#1} (#3) #2}
\def \prl#1#2#3{Phys.\ Rev.\ Lett.\ {\bf#1} (#3) #2}
\def \prp#1#2#3{Phys.\ Rep.\ {\bf#1} (#3) #2}
\def \ptp#1#2#3{Prog.\ Theor.\ Phys.\ {\bf#1} (#3) #2}
\def \ppn#1#2#3{Prog.\ Part.\ Nucl.\ Phys.\ {\bf#1} (#3) #2}
\def \rmp#1#2#3{Rev.\ Mod.\ Phys.\ {\bf#1} (#3) #2}
\def \rp#1{~~~~~\ldots\ldots{\rm rp~}{#1}~~~~~}
\def \si90{25th International Conference on High Energy Physics, Singapore,
Aug. 2-8, 1990}
\def \zpc#1#2#3{Zeit.\ Phys.\ C {\bf#1} (#3) #2}
\def \zpd#1#2#3{Zeit.\ Phys.\ D {\bf#1} (#3) #2}

\newpage

\section*{Figure captions}
\vskip1cm

\hskip0.6cm Figure 1: 
Feynman diagrams for the process
$e^+e^-\to \tau^+\tau^- \nu_e\bar\nu_e$.

Figure 2:
Feynman diagrams for the process
$e^+e^-\to \tau^+\tau^-  \nu_\mu\bar\nu_\mu$.

Figure 3:
Feynman diagrams for the process
$e^+e^-\to \tau^+\tau^- \nu_\tau\bar\nu_\tau $.

Figure 4:
The total cross section $e^+e^-\to \tau^+\tau^- \nu \bar\nu$
dependence on $\Delta a$ (solid line) and $b$ (dashed line).

Figure 5:
Distributions in $T$ correlation, the $\tau$ 
momentum $|\vec{p_{\tau}}|$, the angle between $\tau$ and the beam, 
$\cos \theta_{e \tau}$ and the $\tau$ pair invariant mass, $M_{\tau^+ \tau^-}$.

Figure 6:
Contribution of Higgs diagrams to the differential
distributions in the  $e^+e^-\to \tau^+\tau^- \nu \bar\nu$ process for
standard model ($a=1, b=0$, black dots) and $a=0.5, b=0.5$ (crossed dots).

Figure 7:
Sensitivity functions
$\frac{\sigma^{New} - \sigma^{SM}}{\Delta \sigma^{exp}}$ for each bin of the
distributions, where as an example we took  $a=1, b=0.5$ for new physics.

Figure 8:
Relative contributions to the sensitivity functions
of electron neutrino (black
dots), muon neutrino (rectangular crossed dots), and tau neutrino 
(crossed round dots) channels.

Figure 9:
Allowed regions for ${\cal L} =$ 100 fb$^{-1}$
(the domain inside the solid circle -- there is no inner circle for this
case), 1 ab$^{-1}$ (domain confined by short-dashed contours), and 10 ab$^{-1}$ 
(long-dashed contours).

Figure 10:
The allowed region for independent $\Delta a$ and $b$ parameters 
at 95\% confidence level is the area between the solid circles.
The horizontal band with long-dashed lines is the allowed region for the $b$
parameter keeping $a=1$. The two vertical bands with short-dashed lines are the 
the allowed region for the $\Delta a$
parameter keeping $b=0$

\newpage

\end{document}